\documentclass[conference]{IEEEtran}
\IEEEoverridecommandlockouts
\usepackage[backend=biber,style=ieee,citestyle=numeric-comp,maxbibnames=1]{biblatex}
\usepackage{enumitem}
\usepackage{amsmath,amssymb,amsfonts}
\usepackage{algorithmic}
\usepackage{graphicx}
\usepackage{textcomp}
\usepackage{xcolor}
\def\BibTeX{{\rm B\kern-.05em{\sc i\kern-.025em b}\kern-.08em
    T\kern-.1667em\lower.7ex\hbox{E}\kern-.125emX}}

\usepackage{hyperref}  
\usepackage{orcidlink}
\usepackage{tabularx}
\usepackage{graphicx}
\usepackage{subcaption}
\usepackage{algorithm}
\usepackage{footmisc}

\bibliography{references}

\AtEveryBibitem{
    \clearfield{urlyear}
    \clearfield{urlmonth}
}

\makeatletter
\def\ps@IEEEtitlepagestyle{
  \def\@oddfoot{\mycopyrightnotice}
  \def\@evenfoot{}
}
\def\mycopyrightnotice{
  {\footnotesize © 2025 IEEE. Personal use is permitted. Published in IEEE BigData 2025. DOI: \href{https://doi.org/10.1109/BigData66926.2025.11401516}{10.1109/BigData66926.2025.11401516}\hfill}
  \gdef\mycopyrightnotice{}
}
\makeatother
\begin{document}

\title{A Taxonomy and Resolution Strategy for Client-Level Disagreements in Federated Learning}


\author{\IEEEauthorblockN{Daan Rosendal\IEEEauthorrefmark{1}, Ana Oprescu\IEEEauthorrefmark{2}}
\IEEEauthorblockA{\textit{Complex Cyber Infrastructure, University of Amsterdam, Amsterdam, The Netherlands} \\
\IEEEauthorrefmark{1}\orcidlink{0000-0002-3447-9000}, \href{mailto:daanrosendal@outlook.com}{daanrosendal@outlook.com}, \IEEEauthorrefmark{2}\orcidlink{0000-0001-6376-0750}, \href{mailto:a.m.oprescu@uva.nl}{a.m.oprescu@uva.nl}
}
}

\maketitle

\begin{abstract}
Federated Learning (FL) typically assumes unconditional collaboration, a premise that overlooks the complexities of real-world, multi-stakeholder environments in which clients may need to exclude one another for strategic, regulatory, or competitive reasons. This paper addresses this gap, which we term `client-level disagreements,' by first introducing a taxonomy of such scenarios. We then propose a robust, multi-track resolution strategy that guarantees strict client exclusion by creating and managing isolated model update paths (`tracks'), thereby preventing the cross-contamination and unfairness issues present in naive strategies. Through an empirical evaluation of our custom simulation system across 34 scenarios using the MNIST and N-CMAPSS datasets, we validate that our approach correctly handles permanent, temporal, and overlapping disagreement patterns. Our scalability analysis reveals the server-side resolution algorithm\textquotesingle{}s overhead is negligible (\boldmath$<1$ ms per round) even under heavy load. The primary scalability constraint is the client-side training load from participating in multiple tracks, a cost that we show can be effectively mitigated by a submodel reuse strategy. This work presents a scalable and architecturally sound method for managing client-level disagreements, and enhances the practical applicability of FL in settings where policy compliance and strategic control are non-negotiable.
\end{abstract}


\section{Introduction}
\label{sec:introduction}
Federated Learning (FL) provides an approach for training machine learning models collaboratively across decentralized data sources \cite{McMahan2023FL}. Its ability to maintain data locality makes it highly suitable for applications in sensitive domains like healthcare, finance, and aviation, where privacy regulations and competitive concerns often prevent data centralization.

A limitation in existing FL setups is the implicit assumption of unconditional collaboration, as this premise can fail in practice: \textcolor{black}{competing airlines in a predictive maintenance collaboration may block model contributions to protect insights derived from proprietary data; hospitals may be required by policy to exclude patient data from certain sources; or banks may need to enforce mutual isolation to comply with data sovereignty laws. We formally define such scenarios -- where autonomous clients selectively exclude others for strategic, regulatory, or competitive reasons -- as \textbf{client-level disagreements}.}

While many studies discuss data-level disagreements across clients ~\cite{Ahmed2020Active, rodriguez2024federated, deng2020adaptive, li2020enhancing, shen2023share, yang2023hierarchical, li2022aware, zhang2022finetuning, wang2023hetvis, sattler2020clustered}, primarily arising from non-IID data distributions and system heterogeneity, there is a notable research gap in addressing the higher-level disagreements that emerge between clients themselves. Similarly, current research largely focuses on submodels for optimizing resource efficiency in FL~\cite{alam2022fedrolex, wu2024fiarse, ding2022federated, balakrishnan_diverse_2022, niu_billion-scale_2020, white_evaluating_2024}, overlooking their potential to address client-level disagreements. This highlights another research gap: leveraging submodels for excluding specific clients dynamically during FL. Research on  Clustered FL~\cite{sattler2020clustered, ghosh2022efficient, kim2021dynamic} aims to optimize model performance for heterogeneous data clusters, and does not explicitly apply it to  client-level disagreements. 

Addressing these gaps requires new approaches that integrate submodel generation with flexible client exclusion mechanisms.
We formulate three questions: \textbf{RQ1}) What client-level disagreement scenarios can potentially occur in FL? \textbf{RQ2}) How can client-level disagreements be resolved in FL? and \textbf{RQ3}) How can architectural and algorithmic solutions mitigate the resulting scalability challenges? 

We contribute the following: 1) a formal taxonomy of client-level disagreements, defining a language for expressing exclusion rules and their temporal properties; 2) a robust multi-track resolution algorithm that guarantees strict exclusion and incorporates a submodel reuse mechanism for scalability; 3) an open-source reproducible simulation and validation system; and 4) an empirical validation on two datasets confirming the logical correctness and performance of our algorithm.

\noindent\textbf{Outline.} Sec.~\ref{sec:back+rel-work} summarizes the relevant foundational concepts and reflects on related work. Sec.~\ref{sec:taxonomy} introduces the Federated Learning client-level disagreement taxonomy, and Sec.~\ref{sec:resolution_strategy} presents our robust resolution strategy. Sec.~\ref{sec:system_design} describes our system design and implementation. We present the experimental setup in Sec.~\ref{sec:experimental_setup} and collect our results in Sec.~\ref{sec:results}. We answer the research questions and reflect on our findings in Sec.~\ref{sec:discussion} and conclude in Sec.~\ref{sec:conclusion}.

\section{Background and related work} 
\label{sec:back+rel-work}
Dalgkitsis et al.~\cite{dalgkitsis2024secure} validate the enforcement of high-level disagreements in a distributed FL system, yet model them as a binary `full exclusion' where clients either participate or do not. In contrast, our work introduces a granular taxonomy of disagreements, \textit{e.g.}, inbound, outbound, temporal, and proposes a multi-track architectural strategy required to resolve the resulting complexity.

\textcolor{black}{Clustered FL (CFL)~\cite{sattler2020clustered, kim2021dynamic} is not a direct baseline as it automatically groups clients by data heterogeneity to optimize accuracy, whereas our system enforces explicit, policy-based disagreements.} \textcolor{black}{Personalized FL methods, \textit{e.g.}, pFedMe~\cite{dinh2022personalizedfederatedlearningmoreau}, Per-FedAvg~\cite{fallah2020personalizedfederatedlearningmetalearning} assume all clients contribute to a central model before specialization and offer no exclusion mechanism. In contrast, our work defines the scope of collaboration itself.}

\vspace{-2mm}
\section{Disagreement Taxonomy}
\label{sec:taxonomy}
To address client-level disagreements, we define a formal taxonomy categorizing how clients may deviate from unconditional collaboration. This taxonomy provides a structured language for defining exclusion policies based on their impact on communication, data, and temporal properties. Table~\ref{tab:disagreement-types-overview} lists the primary disagreement types and their exclusion policies.

\begin{table}[htbp]
    \centering
    \caption{Overview of Disagreement Types}\vspace{-2mm}
    \label{tab:disagreement-types-overview}
    \begin{tabular}{|p{1.3cm}|p{6.7cm}|}
        \hline
        \textbf{Type} & \textbf{Description of disagreement scenario} \\
        \hline\hline
        \multicolumn{2}{|l|}{\textbf{Communication-Based Exclusions}} \\
        \hline
        Full  & Client completely withdraws from the FL system. \\
        \hline
        Inbound  & Client rejects model updates from specific participants. \\
        \hline
        Outbound  & Client prevents its model from contributing to others\textquotesingle{} global models. \\
        \hline
        Bidirectional  & Mutual exclusion preventing any model influence between clients. \\
        \hline
        \multicolumn{2}{|l|}{\textbf{Data-Based Exclusions}} \\
        \hline
        Partial data & Client restricts the use of parts of its local data for training. \\
        \hline
        \multicolumn{2}{|l|}{\textbf{Temporal Properties}} \\
        \hline
        Temporary & An exclusion automatically expiring after a defined period. \\
        \hline
        Deep (Retrospective) & An exclusion that retroactively removes historical model contributions. \\
        \hline
    \end{tabular}
    \vspace{-3mm}
\end{table}

\begin{figure}[htbp]
    \centering
    \includegraphics[width=\columnwidth]{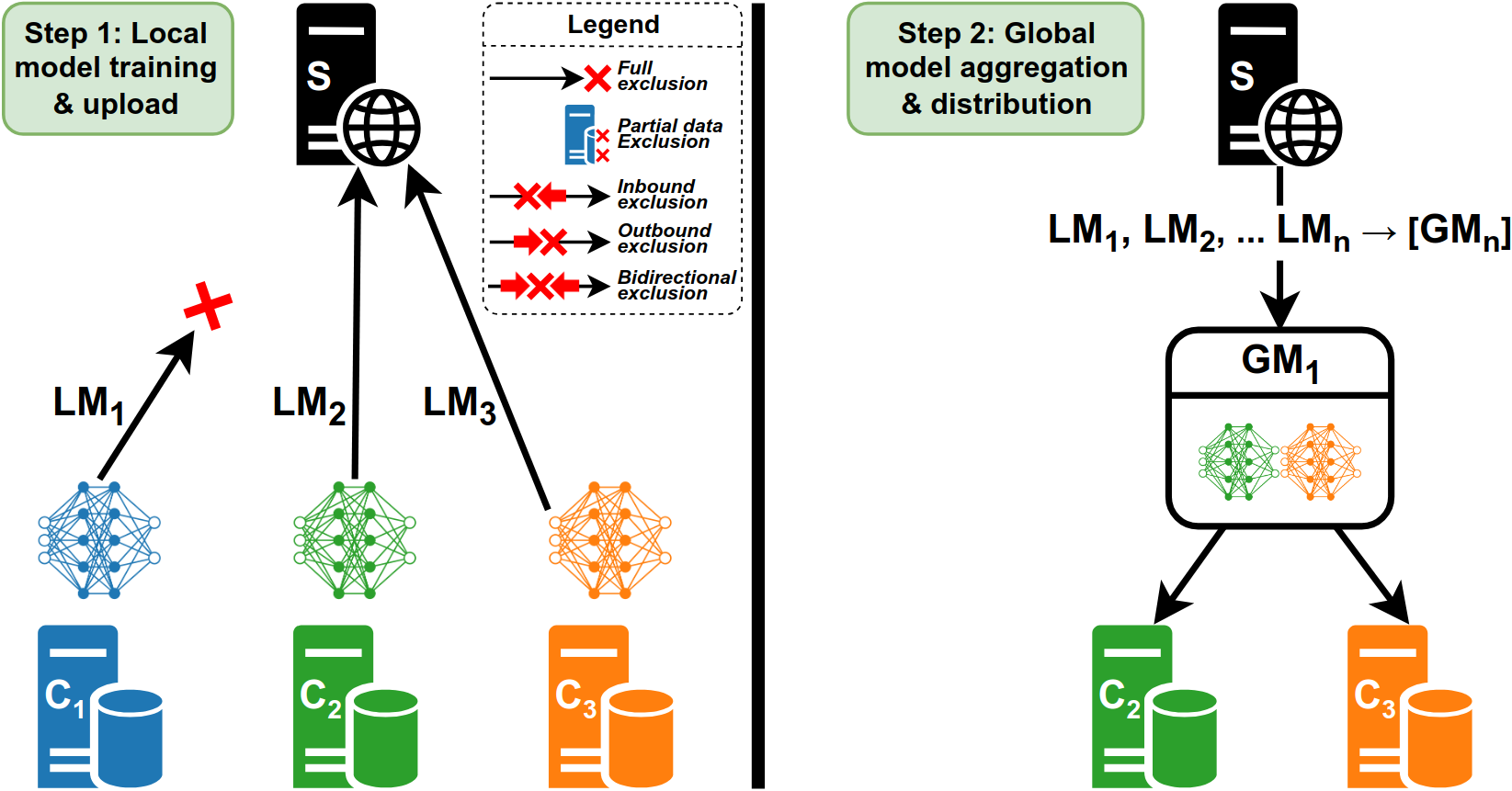}\vspace{-2mm}
    \caption{Full exclusion: client C\textsubscript{1} fully removes itself from FL. \vspace{0.1mm}}
    \label{fig:full-exclusion}
\end{figure}

\noindent\textit{A. Exclusion Types}\\
\noindent\textbf{Full Exclusion.}
The most basic form of disagreement is a full exclusion, where a client completely withdraws from the FL process, as shown in Fig.~\ref{fig:full-exclusion}. This entails ceasing to submit local models and no longer receiving any global model updates.

\begin{figure}
    \centering
    \includegraphics[width=\columnwidth]{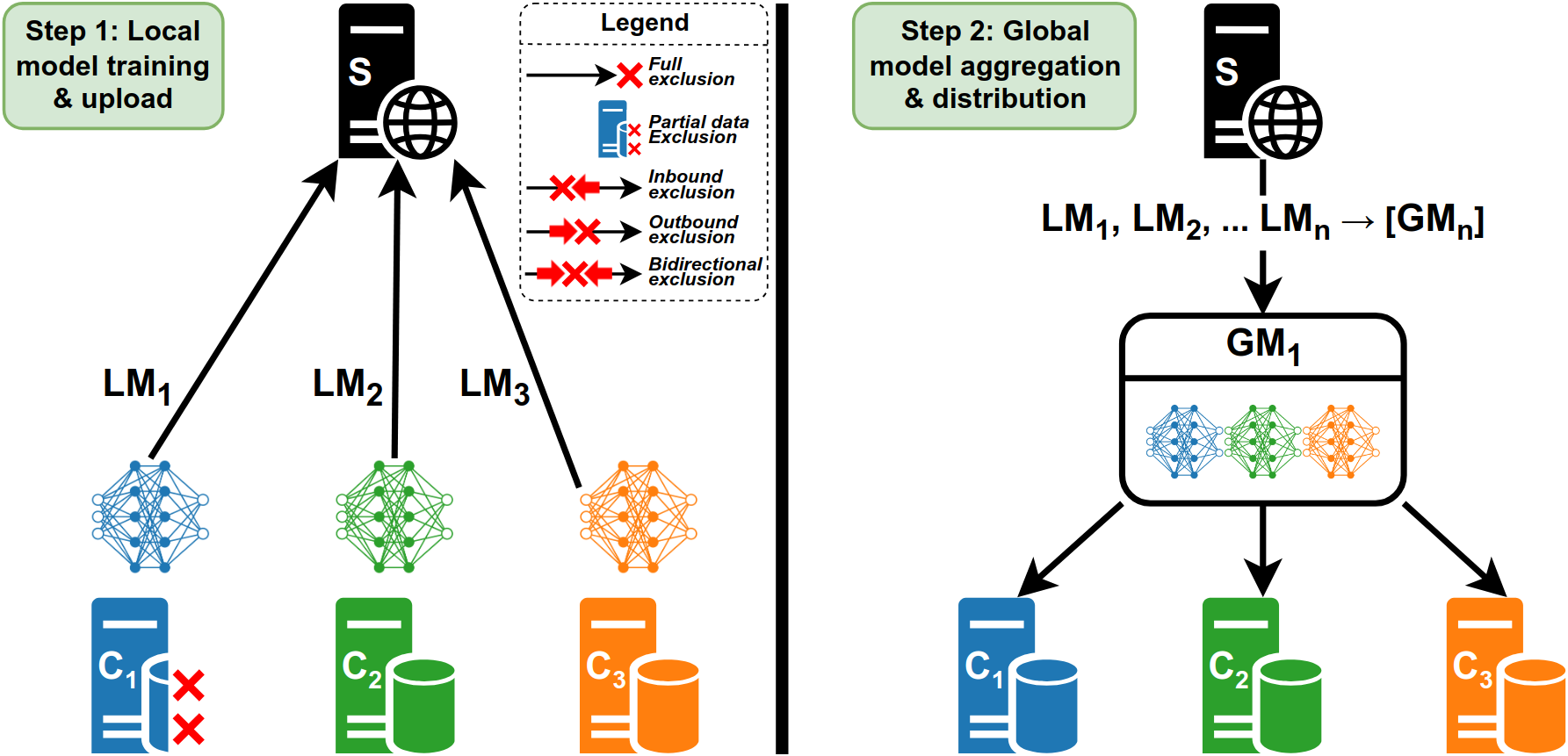}\vspace{-2mm}
    \caption{Partial data exclusion, where C\textsubscript{1} withholds a subset of its local data but continues to participate in model aggregation.}
    \label{fig:partial-data-exclusion}
\end{figure}

\vspace{-7mm}
\noindent\textbf{Partial Data Exclusion.}
Fig.~\ref{fig:partial-data-exclusion} shows a client withholding a specific subset of its local data from the training process. This gives fine-grained control, enabling clients to protect sensitive data segments and still contribute to the collaborative model.

\begin{figure}
    \centering
    \includegraphics[width=\columnwidth]{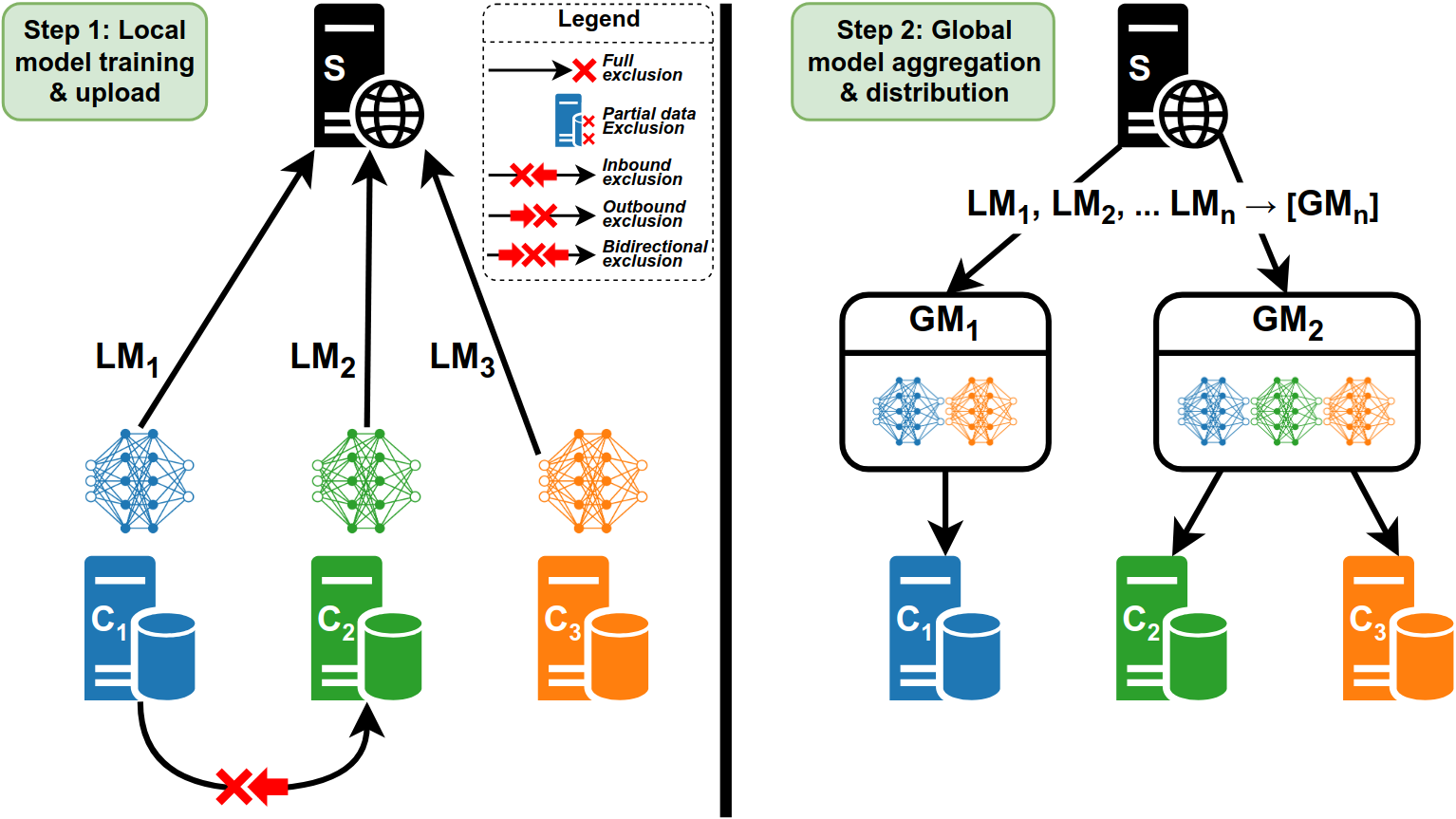}\vspace{-2mm}
    \caption{Inbound exclusion, where C\textsubscript{1} requires a personalized global model (GM\textsubscript{1}) that omits C\textsubscript{2}\textquotesingle{}s contribution. \vspace{-3mm}}
    \label{fig:inbound-one-excludes-one}
     \vspace{-3mm}
\end{figure}

\noindent\textbf{Inbound Exclusion} occurs when a client refuses to incorporate the model updates from one or more specific participants into its own version of the global model. Fig.~\ref{fig:inbound-one-excludes-one} shows a case where C\textsubscript{1} inbound excludes C\textsubscript{2}: the server generates a personalized submodel (GM\textsubscript{1}) for C\textsubscript{1} that omits C\textsubscript{2}\textquotesingle{}s contribution, while other clients receive the standard global model (GM\textsubscript{2}).

\begin{figure}
    \centering
    \includegraphics[width=1\columnwidth]{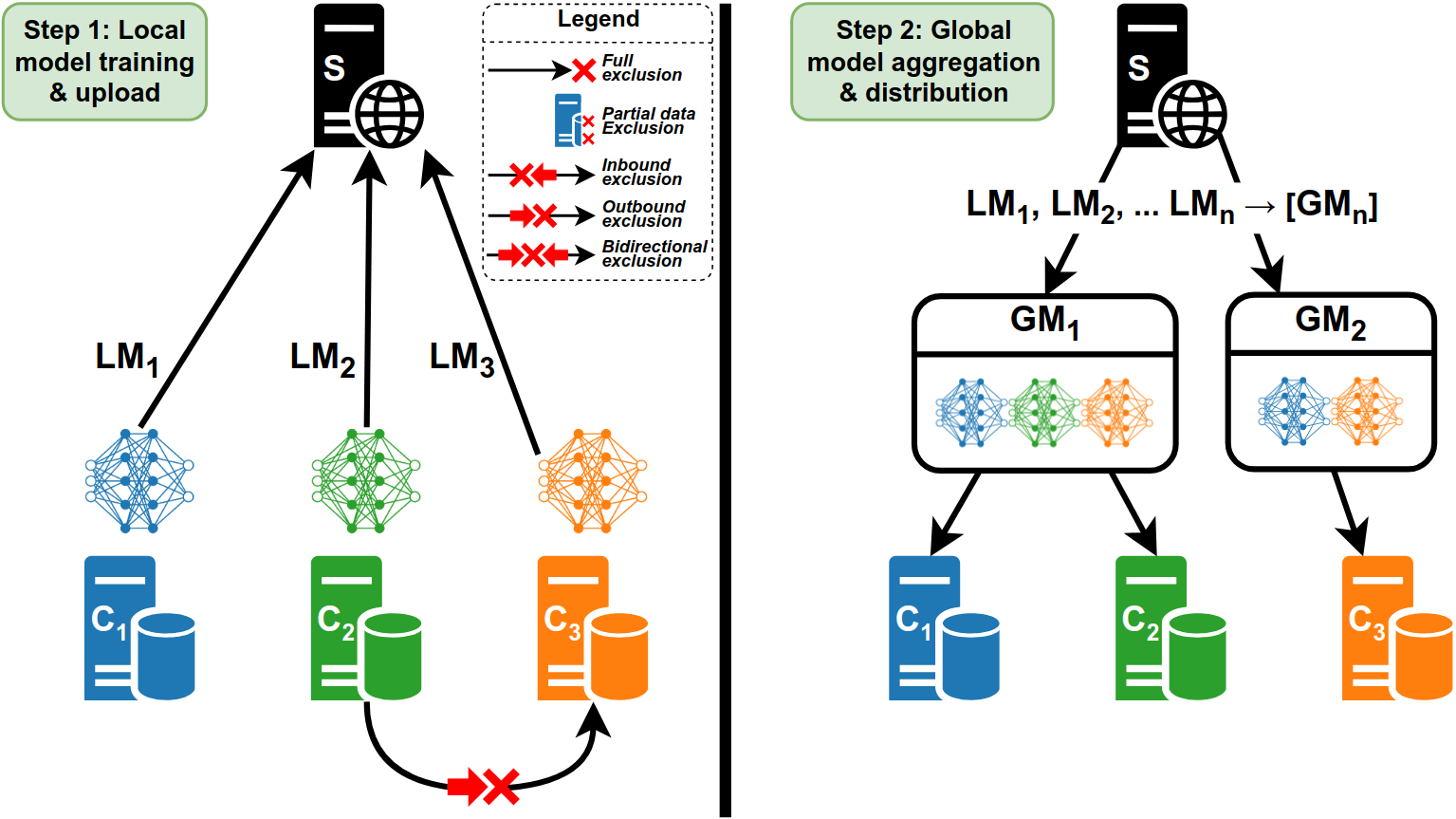}\vspace{-2mm}
    \caption{Outbound exclusion, where C\textsubscript{2} prevents its model from being incorporated into C\textsubscript{3}\textquotesingle{}s global model.}
    \label{fig:outbound-one-excludes-one}
\end{figure}

\noindent\textbf{Outbound Exclusion.} As the inverse of an inbound exclusion, it allows a client to prevent its local model from being included in the global models of specific other participants. Fig.~\ref{fig:outbound-one-excludes-one} shows a case where C\textsubscript{2} prevents its model from being used in C\textsubscript{3}\textquotesingle{}s global model. Consequently, the server must generate a separate model (GM\textsubscript{2}) for C\textsubscript{3} that excludes C\textsubscript{2}\textquotesingle{}s update, while the other clients receive the standard global model (GM\textsubscript{1}).

\begin{figure}
    \centering
    \includegraphics[width=1\columnwidth]{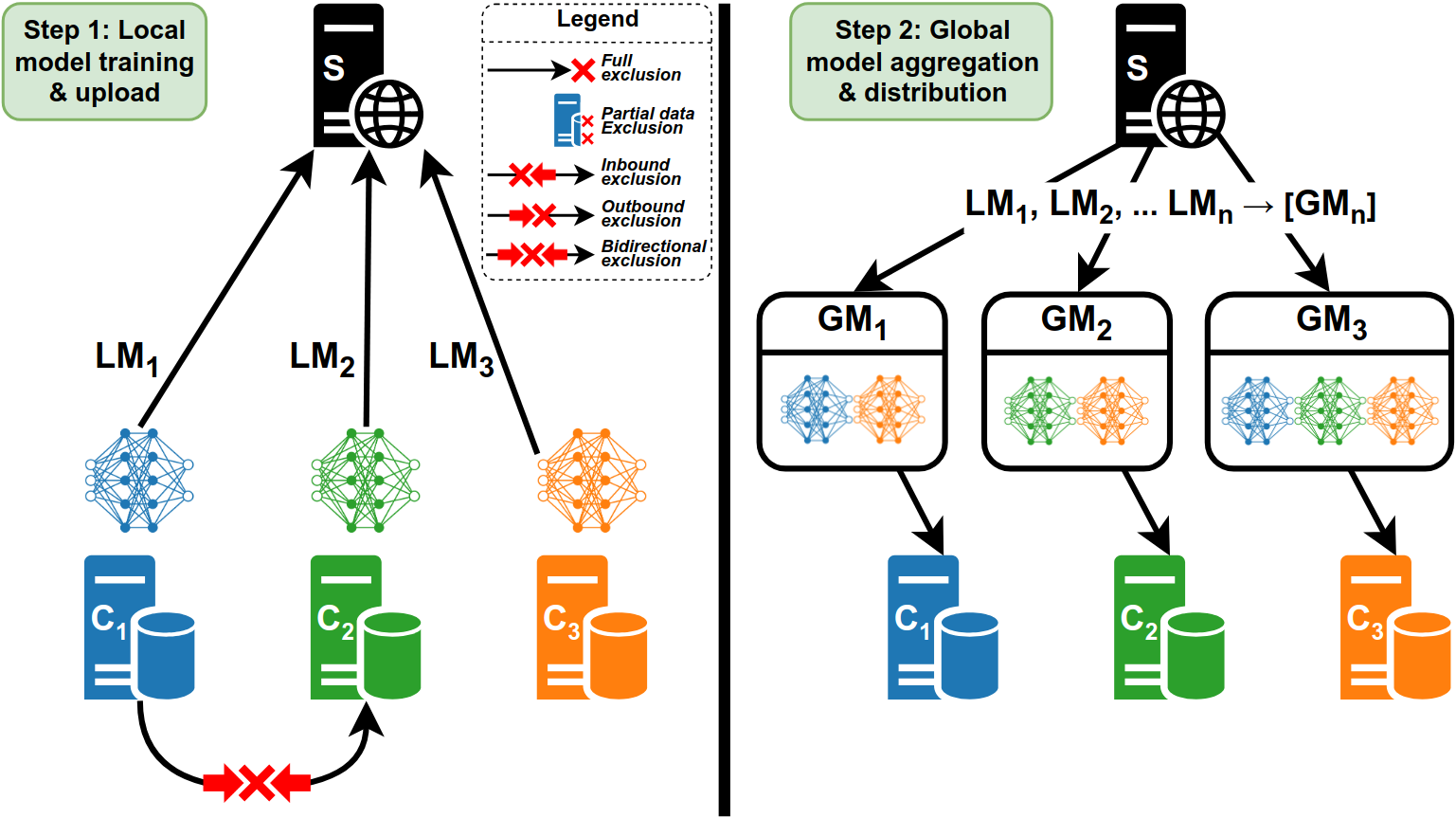}\vspace{-2mm}
    \caption{Bidirectional exclusion (one excludes one), establishing mutual isolation between C\textsubscript{1} and C\textsubscript{2}.}
    \label{fig:bidirectional-one-excludes-one}
    \vspace{-3mm}
\end{figure}

\noindent\textbf{Bidirectional Exclusion} is a composite disagreement combining inbound and outbound rules, where two clients mutually prevent any model influence from being exchanged between them. This establishes complete training isolation. Fig.~\ref{fig:bidirectional-one-excludes-one} illustrates a case where C\textsubscript{1} imposes a bidirectional exclusion on C\textsubscript{2}. The server must now generate three distinct models: GM\textsubscript{1} for C\textsubscript{1} (excluding C\textsubscript{2}), GM\textsubscript{2} for C\textsubscript{2} (excluding C\textsubscript{1}), and GM\textsubscript{3} for C\textsubscript{3}, which contains all updates.

\noindent\textit{B. Temporal Dynamics}\\
The exclusion types can be further modified by temporal properties that define their lifecycle and historical impact.

\noindent\textbf{Exclusion Duration.} Disagreements are not necessarily permanent. An exclusion can be defined as \textit{indefinite} (persisting until manually revoked) or \textit{temporary} (automatically expiring after a predefined number of rounds). This enables dynamic collaboration patterns, such as data embargoes.

\noindent\textbf{Exclusion Depth} defines whether an exclusion applies only prospectively or retrospectively. A \textit{shallow} exclusion affects only future model aggregations, while a \textit{deep} exclusion, signified by a deep icon (downward arrow over waves), retroactively removes the historical influence of an excluded client\textquotesingle{}s past contributions. Deep exclusions are significantly more complex, requiring mechanisms such as federated unlearning~\cite{Gao2024VeriFi, Liu2024Survey} or historical recalculation.

\begin{figure}
    \centering
    \includegraphics[width=1\columnwidth]{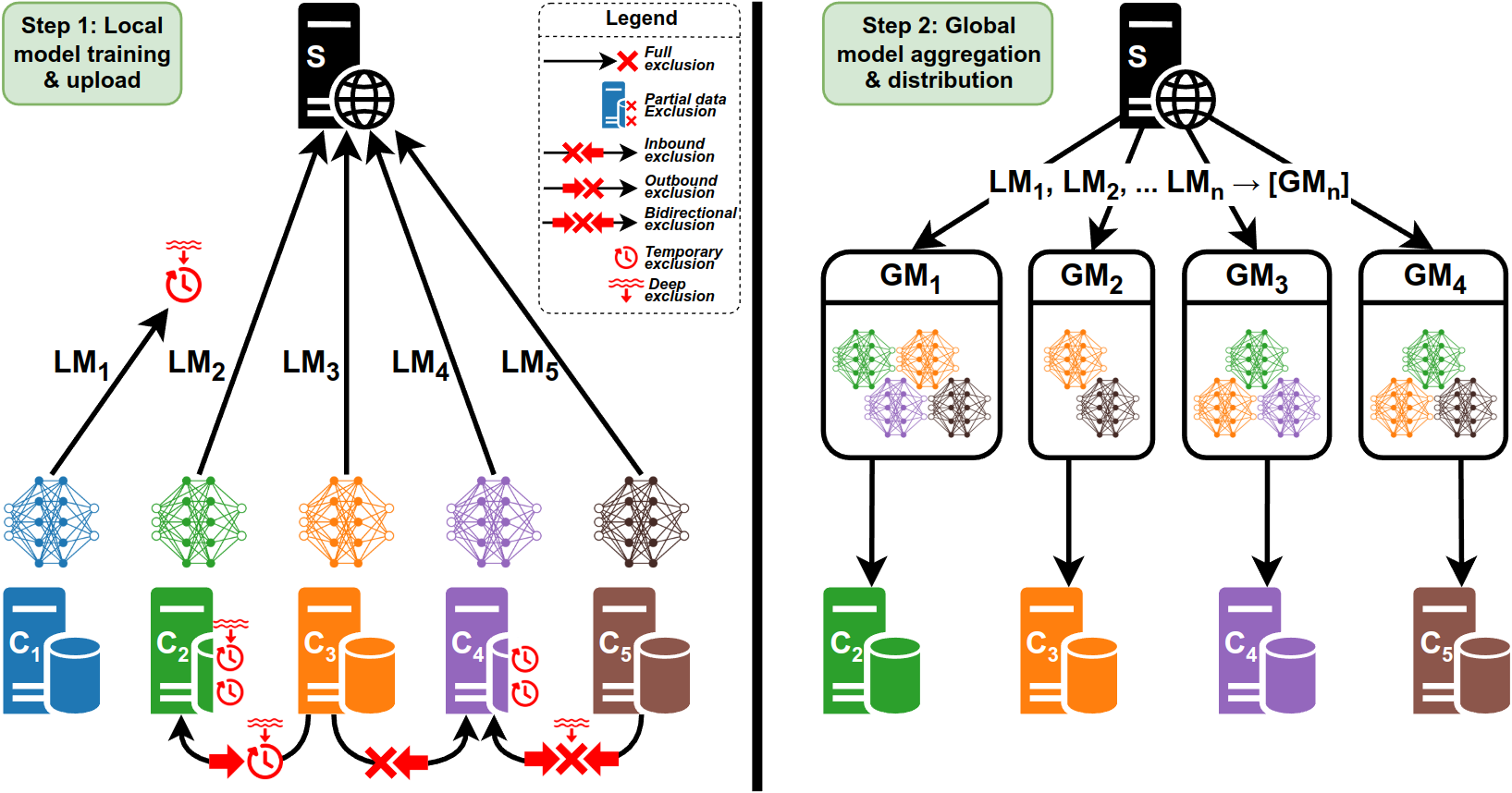}\vspace{-2mm}
    \caption{Multiple exclusion types disagreement scenario with temporal modifiers for duration (clock) and depth (deep icon).}
    \label{fig:combination-temporary-depth}
    \vspace{-5mm}
\end{figure}

Fig.~\ref{fig:combination-temporary-depth} shows a complex scenario combining multiple exclusion types with  temporal modifiers. Client C\textsubscript{1} enforces a temporary, deep full exclusion, aiming to retroactively remove its historical contributions for a limited period. Client C\textsubscript{2} enacts a similar temporary, deep partial data exclusion. Client C\textsubscript{3} initiates two separate exclusions: a temporary, deep inbound exclusion against C\textsubscript{2} and an indefinite inbound exclusion against C\textsubscript{4}. Concurrently, client C\textsubscript{4} applies an indefinite partial data exclusion, while clients C\textsubscript{4} and C\textsubscript{5} engage in an indefinite, deep bidirectional exclusion, establishing complete and mutual isolation. Earlier examples focus on isolated disagreements, while this scenario shows how the complexity escalates rapidly when multiple exclusion rules are layered together.

\section{Resolution Strategy}
\label{sec:resolution_strategy}
\vspace{-1mm}

Resolving the client-level disagreements defined in Sec.~\ref{sec:taxonomy} requires an architectural strategy that can enforce client-specified exclusions without compromising the integrity and fairness of the collaborative learning process. 

\begin{figure}
    \centering
    \includegraphics[width=\columnwidth]{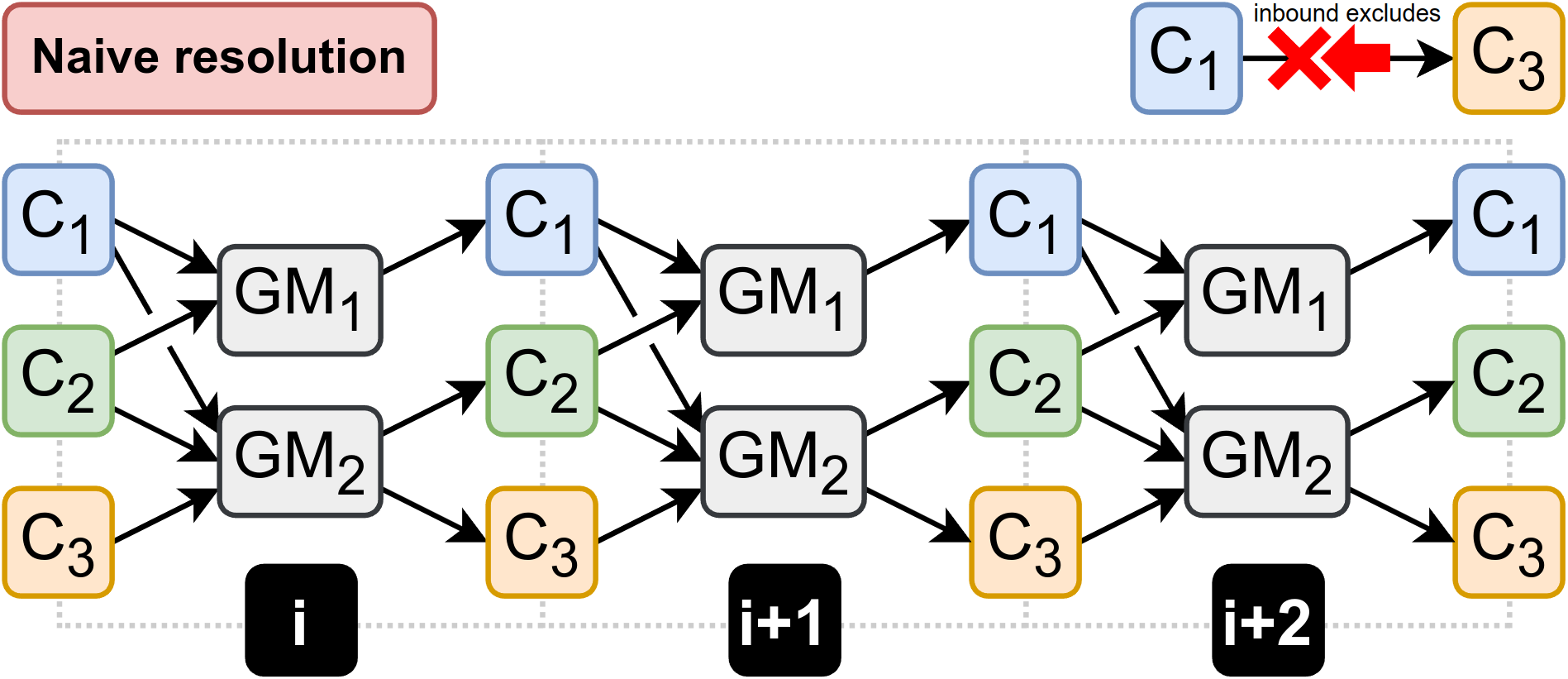}\vspace{-2mm}
    \caption{The naive resolution strategy. C\textsubscript{1} inbound excludes C\textsubscript{3}, receiving a personalized model (GM\textsubscript{1}). However, intermediary clients like C\textsubscript{2} still aggregate all updates.}
    \label{fig:naive}
\end{figure}

\noindent\textit{A. The Naive Approach and its Flaws}\\
A simple disagreement resolution approach is to generate a personalized global model for each client based on its specific exclusion rules for the current round. Fig.~\ref{fig:naive} shows how, if client C\textsubscript{1} inbound excludes C\textsubscript{3}, the server aggregates models from C\textsubscript{1} and C\textsubscript{2} to create C\textsubscript{1}\textquotesingle{}s new global model. Meanwhile, C\textsubscript{2} and C\textsubscript{3} might receive a global model aggregating all clients.

Although simple, this strategy fails over multiple rounds due to two critical issues: cross-contamination and unfairness.

\noindent\textbf{Cross-Contamination.}
Despite a direct exclusion in place, the influence of an excluded client can indirectly propagate to the excluding client through an intermediary. Fig.~\ref{fig:cross-contamination} shows C\textsubscript{1} excluding C\textsubscript{3}, while C\textsubscript{2} continues to incorporate updates from C\textsubscript{3}. In the next round (\textit{i}+1), when C\textsubscript{1} aggregates its model with C\textsubscript{2}\textquotesingle{}s, it inadvertently re-introduces the influence of C\textsubscript{3}. This leakage violates the principle of strict exclusion.

\begin{figure}
    \centering
    \includegraphics[width=\columnwidth]{figures/resolutions/cross-contamination.png}\vspace{-2mm}
    \caption{Cross-contamination in the naive approach. The influence of the excluded client (C\textsubscript{3}) leaks back to C\textsubscript{1} via the intermediary client C\textsubscript{2} in subsequent rounds.\vspace{-4mm}}
    \label{fig:cross-contamination}
\end{figure}

\begin{figure}
    \centering
    \includegraphics[width=\columnwidth]{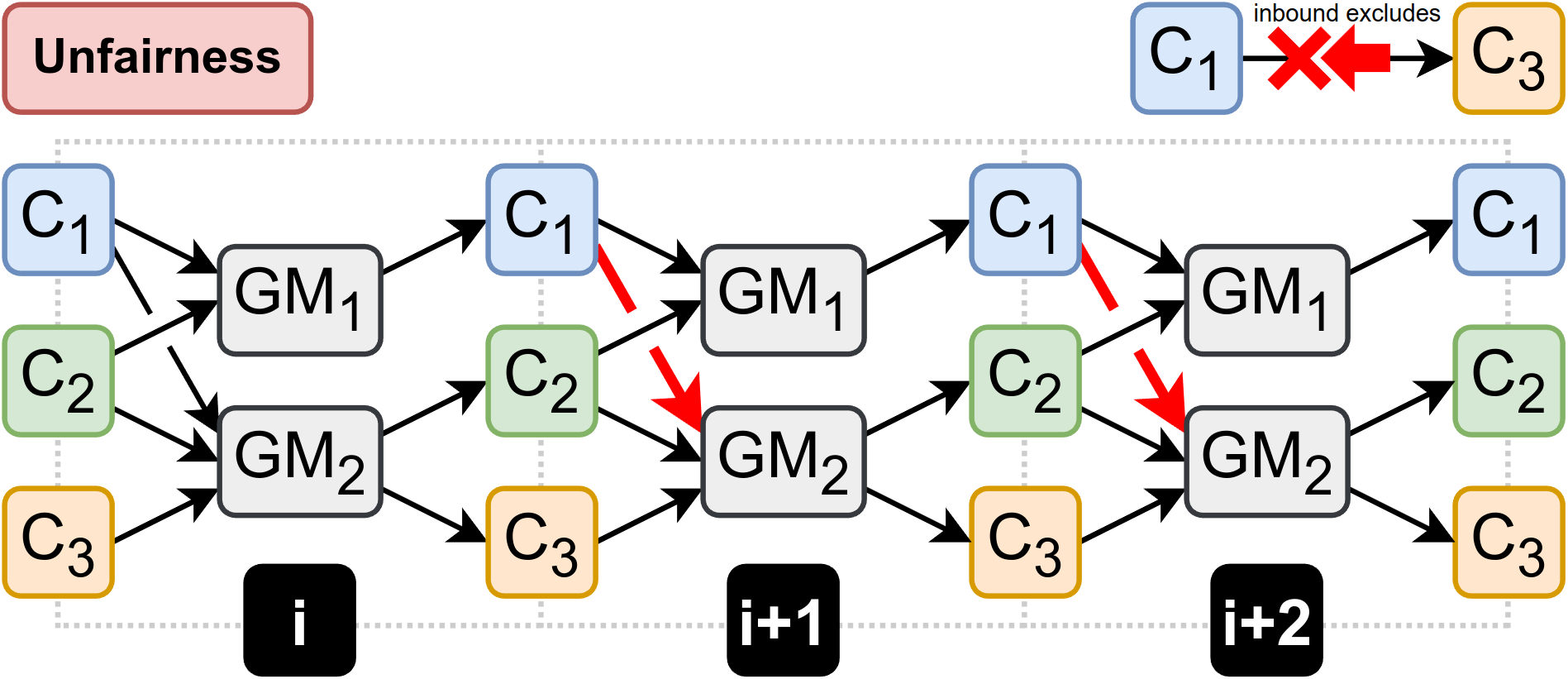}\vspace{-2mm}
    \caption{Unfairness in the naive approach. C\textsubscript{1} contributes to C\textsubscript{3}\textquotesingle{}s model, but its updates are based on a personalized model that excluded C\textsubscript{3}, creating an inequitable aggregation.}
    \label{fig:unfairness}
\end{figure}

\noindent\textbf{Unfairness.}
This approach unfairly imbalances contributions. Fig.~\ref{fig:unfairness} shows C\textsubscript{1} training on a model that excludes C\textsubscript{3}, while C\textsubscript{1}\textquotesingle{}s updates contribute to the global model C\textsubscript{3} receives. Thus C\textsubscript{3} is influenced by a model trained under different conditions, \textit{i.e.}, from C\textsubscript{1}\textquotesingle{}s perspective, without C\textsubscript{3}. This undermines the principle of equitable participation where all contributions are generated under consistent training assumptions.

\vspace{4mm}

\noindent\textit{B. The Robust Multi-Track Approach}\\
To overcome these limitations, we propose a robust resolution strategy that guarantees strict exclusion and fairness by creating and managing independent model update paths, i.e., \textit{tracks}. Each unique set of client disagreements results in a distinct track, ensuring completely isolated model updates.

\begin{figure}
\vspace{-2mm}
    \centering
    \includegraphics[width=\columnwidth]{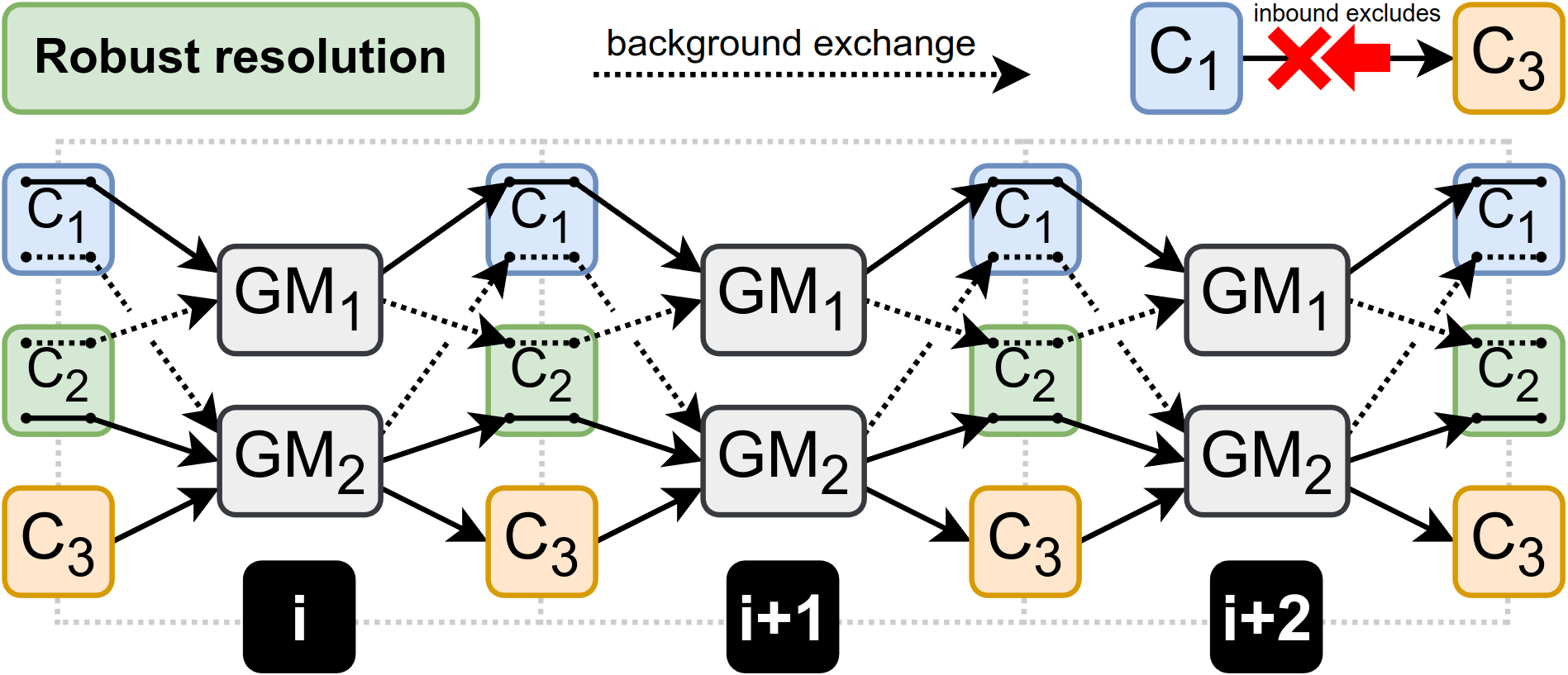}\vspace{-2mm}
    \caption{The robust multi-track approach. Two separate and isolated tracks (GM\textsubscript{1} and GM\textsubscript{2}) are created to serve the different disagreement requirements of the clients.\vspace{-3mm}}
    \label{fig:robust}
\end{figure}

Fig.~\ref{fig:robust} shows how the system creates two separate tracks when C\textsubscript{1} excludes C\textsubscript{3}. Track GM\textsubscript{1} is the primary track for C\textsubscript{1}, aggregating only from C\textsubscript{1} and C\textsubscript{2}. Track GM\textsubscript{2} is the primary track for C\textsubscript{2} and C\textsubscript{3}, aggregating from all clients.

Crucially, to ensure all tracks evolve correctly, clients may need to perform \textit{background participation}. For example, although C\textsubscript{1}\textquotesingle{}s primary model is GM\textsubscript{1}, it must also train on GM\textsubscript{2} and submit an update for that track separately to ensure C\textsubscript{2} and C\textsubscript{3} receive a fairly aggregated model. This mechanism solves the core problems of the naive approach.

\begin{figure}
    \centering
    \includegraphics[width=\columnwidth]{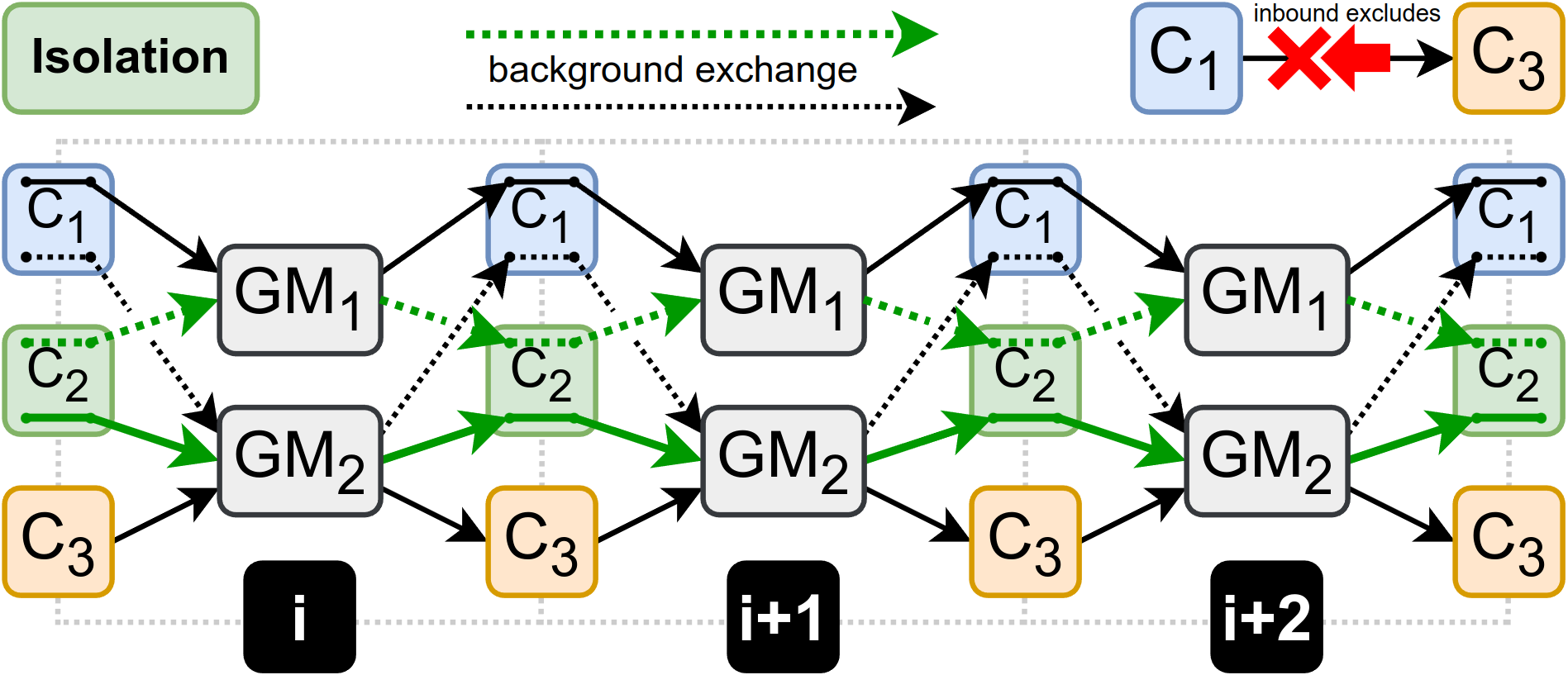}\vspace{-2mm}
    \caption{Isolation in the robust approach. Updates are track-specific, preventing the influence of an excluded client (C\textsubscript{3}) from propagating across tracks.}
    \label{fig:isolation}
\end{figure}

\noindent\textbf{Isolation.}
The multi-track architecture enforces strict isolation, preventing cross-contamination. Fig.~\ref{fig:isolation} shows how client C\textsubscript{2} participates in both tracks by generating a distinct model update for each track based on its respective specific model state. The update for GM\textsubscript{1} is computed without any influence from C\textsubscript{3}, effectively blocking the indirect contamination pathway.

\begin{figure}
    \centering
    \includegraphics[width=\columnwidth]{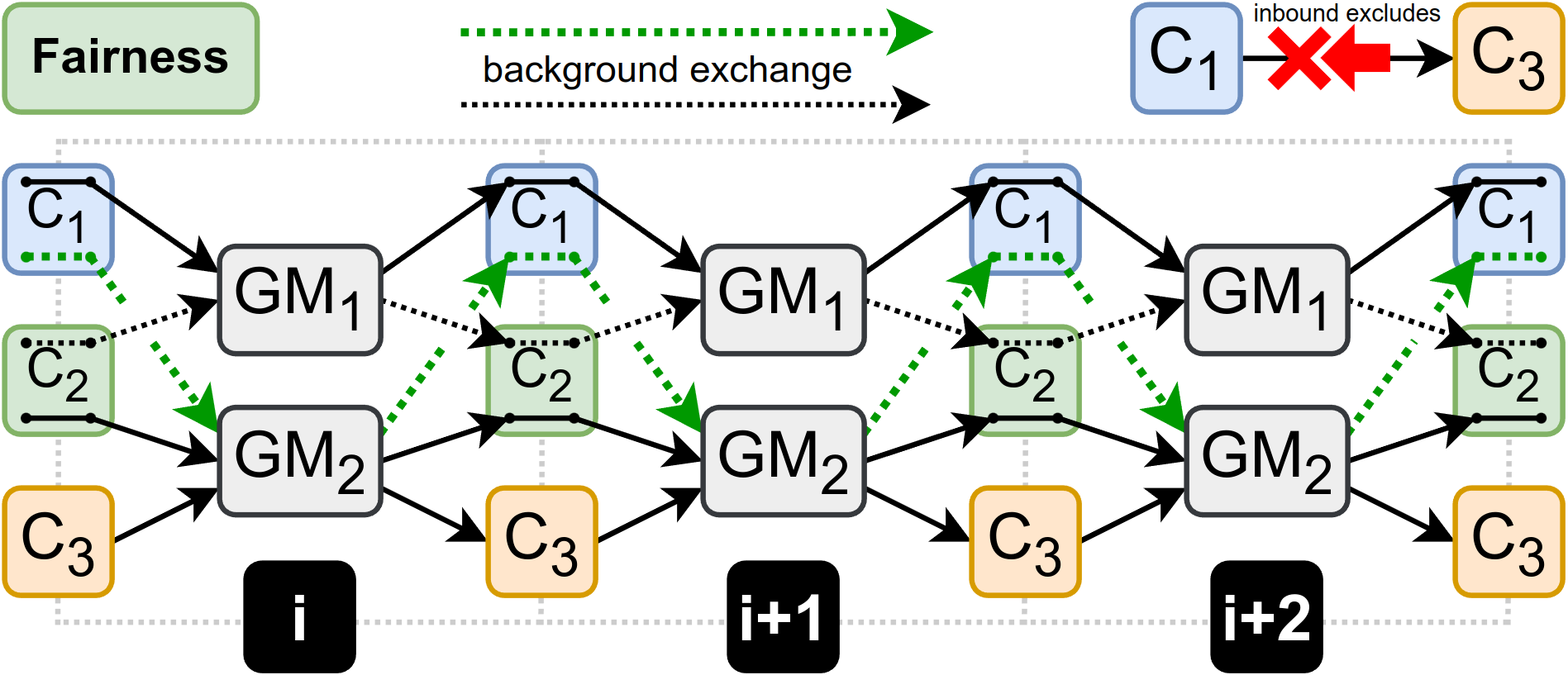}\vspace{-2mm}
    \caption{Fairness in the robust approach. Through background participation, C\textsubscript{1}\textquotesingle{}s contribution to C\textsubscript{3}\textquotesingle{}s track (GM\textsubscript{2}) is generated under fair and consistent conditions.\vspace{-3mm}}
    \label{fig:fairness}
\end{figure}

\noindent\textbf{Fairness.}
Our robust approach also ensures fairness. Fig.~\ref{fig:fairness} shows C\textsubscript{1} performing a background participation for track GM\textsubscript{2}. This means its contribution to the model C\textsubscript{3} receives is based on training GM\textsubscript{2}, which reflects the full set of participants. Thus, C\textsubscript{3} receives a model aggregated from contributions generated under consistent, track-specific conditions.

\section{System Design and Implementation}
\label{sec:system_design}

To empirically validate our robust multi-track resolution strategy, we design and implement a simulation system\footnote{All source code, tooling, and experimental configurations are publicly available at \url{https://github.com/DaanRosendal/fl-disagreement-resolution}~\cite{rosendal2025fldisagreementresolution}}. The architecture borrows from DYNAMOS, a microservices-based middleware for data exchanges~\cite{stutterheim_2024_dynamos}, its core principles of modularity and policy-driven control. We use a simulation instead of a full distributed deployment to focus on the correctness and performance of the core disagreement resolution logic and facilitate rapid, reproducible experimentation.

\begin{figure*}[htbp]
    \centering
    \begin{subfigure}[b]{0.45\textwidth}
        \centering
        \includegraphics[width=\textwidth]{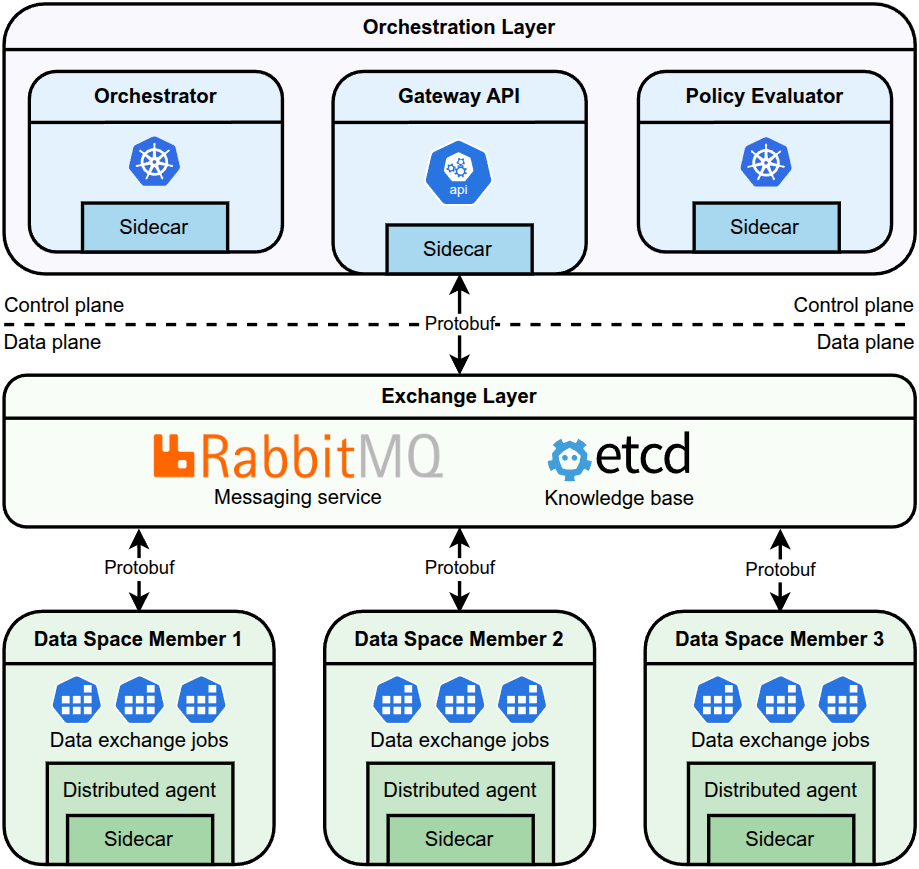} 
        \caption{DYNAMOS abstract design, adapted from \cite{stutterheim_2024_dynamos}.}
        \label{fig:dynamos-arch}
    \end{subfigure}
    \hfill
    \begin{subfigure}[b]{0.5\textwidth}
        \centering
        \includegraphics[width=\textwidth]{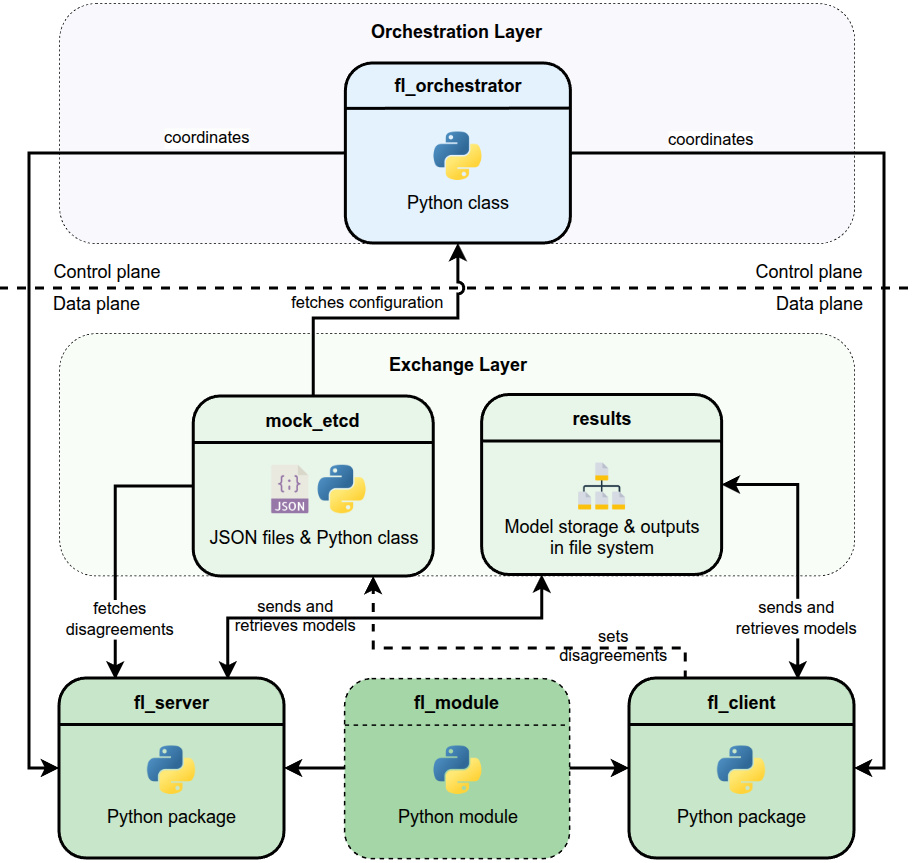}
        \caption{Our DYNAMOS-inspired FL architecture.}
        \label{fig:our-arch}
    \end{subfigure}\vspace{-2mm}
    \caption{A comparison of the original DYNAMOS system and our proof-of-concept FL simulation system.}
    \label{fig:architectures}
    \vspace{-5mm}
\end{figure*}

Fig.~\ref{fig:dynamos-arch} and~\ref{fig:our-arch} show our system\textquotesingle{}s architecture, simulating a distributed FL environment by abstracting the key components and interactions of a DYNAMOS-like system. Our system has three main components: an Orchestrator, an FL Server, and multiple FL Clients. The Orchestrator manages the high-level simulation lifecycle, 
while the core logic resides in the FL Server, emulating DYNAMOS\textquotesingle{}s policy-driven control. 
Based on input policies, the server
executes the disagreement resolution algorithm, determining the required set of isolated model update paths (`tracks'), managing their lifecycle, and performing track-specific aggregation.

Each client is represented by a FL Client instance,
initialized with a private data partition, responsible for local model training, and able to perform multi-track training. Based on server instructions, a client trains not only on its `primary' track to receive its personalized model but also performs background training for other tracks as necessary to ensure system-wide fairness. We simulate the model exchange layer via a shared file system where the server persists aggregated track models and clients retrieve them for local training.

\noindent\textbf{Operational Flow.} The system operates in an iterative, round-based manner. Each FL round consists of four distinct phases, ensuring disagreements are handled dynamically:

\noindent\textbf{1. Analyse}: The FL server begins each round by discovering all active disagreements from the configuration. It then determines the required set of isolated model update paths and creates the necessary model tracks for the current round.

\noindent\textbf{2. Train}: Each FL client loads the appropriate model(s) for its primary and any required background tracks. The client then performs track-based local training on its data and saves the updated models for the server to collect.

\noindent\textbf{3. Aggregate}: The FL server collects the updated models from all participating clients. It aggregates them strictly by track, ensuring that contributions remain isolated within their intended group, and saves these newly aggregated track models.

\noindent\textbf{4. Evaluate}: Finally, the server assesses the training and model performance for each track, measures general system overhead, and logs metadata on track evolution. This cycle repeats until the configured number of rounds is complete.


\noindent\textbf{Core Algorithm.}
The server starts each round with Alg.~\ref{alg:track_creation}. This algorithm translates the high-level disagreement specifications into a concrete set of model tracks. It parses all active disagreements, converting outbound and bidirectional disagreements into an inbound-centric view for each affected client. It then groups clients by their identical exclusion patterns. For each unique pattern, a distinct model track is created. This process is optimized by a consolidation step (submodel reuse) that merges any tracks that, despite originating from different exclusions, result in the exact same set of client members. The output is a definitive map of tracks and the assignment of a primary track to every participating client.

\begin{algorithm}[htbp]
\caption{Disagreement Resolution and Track Creation}
\label{alg:track_creation}
\begin{algorithmic}[1]
\STATE \textbf{Input:} Active Disagreements $D$, All Client IDs $C_{all}$
\STATE \textbf{Output:} Tracks $T_{map}$, Primary Tracks $P_{map}$
\STATE Initialize $exclusion\_map \gets \text{client map to excluded set}$
\FOR{each disagreement $d \in D$}
    \IF{$d.type = \text{"full"}$}
        \STATE Add $d.client$ to $fully\_excluded$
    \ELSIF{$d.type = \text{"inbound"}$}
        \STATE Add $d.target$ to $exclusion\_map[d.client]$
    \ELSIF{$d.type = \text{"outbound"}$}
        \STATE Add $d.client$ to $exclusion\_map[d.target]$
    \ELSIF{$d.type = \text{"bidirectional"}$}
        \STATE Add $d.target$ to $exclusion\_map[d.client]$
        \STATE Add $d.client$ to $exclusion\_map[d.target]$
    \ENDIF
\ENDFOR
\STATE $pattern\_map \gets \text{map of exclusion pattern to clients}$
\FOR{each client $c \in C_{all}$ not in $fully\_excluded$}
    \STATE $pattern \gets \text{sorted tuple of } exclusion\_map[c]$
    \STATE Add $c$ to $pattern\_map[pattern]$
\ENDFOR
\STATE Initialize $T_{map} \gets \emptyset$, $P_{map} \gets \emptyset$
\FOR{each pattern $p$, client group $G \in pattern\_map$}
    \STATE $track\_name \gets \text{generate\_track\_name}(G, p)$
    \STATE $track\_members \gets C_{all} \setminus (p \cup fully\_excluded)$
    \STATE $T_{map}[track\_name] \gets track\_members$
    \FOR{client $c \in G$}
        \STATE $P_{map}[c] \gets track\_name$
    \ENDFOR
\ENDFOR
\STATE \textit{// Add default global track for clients without exclusions}
\STATE Consolidate redundant tracks in $T_{map}$ (Submodel Reuse)
\STATE \textbf{return} $T_{map}$, $P_{map}$
\end{algorithmic}
\end{algorithm}

The \textit{time complexity} of Alg.~\ref{alg:track_creation} is mainly driven by the process of grouping clients based on their exclusion patterns. In the worst-case scenario, where each of the $C$ clients has a dense and unique list of exclusions, the algorithm\textquotesingle{}s complexity is $O(D_{active} + C^2 \log C)$. This is dominated by the need to sort the exclusion list of each client (an operation of up to $O(C \log C)$ per client). In contrast, the average-case performance is significantly more efficient, approaching $O(D_{\text{active}} + C)$ under sparse client exclusion lists and limited number of unique disagreement patterns -- the cost of sorting each exclusion list becomes negligible compared to the linear cost of iterating through the clients. The \textit{space complexity} is determined by the intermediate data structures that map clients to their exclusion patterns. Worst case is $O(C^2)$ -- every client has a unique and dense set of exclusions, requiring the storage of a distinct exclusion list of size up to $C-1$ for each of the $C$ clients in the `exclusion\_map'. The final output data structures also contribute, yet their size is bounded by this initial mapping, making the storage of exclusion patterns the dominant factor in memory usage.

\vspace{-1mm}
\section{Experimental Setup}
\label{sec:experimental_setup}
\vspace{-1.5mm}

We design experiments to: 1) validate the logical correctness of the robust multi-track mechanism in handling diverse and dynamic disagreement scenarios; 2) assess the system\textquotesingle{}s scalability under increasing client load and disagreement complexity; and 3) quantify the computational overhead of the resolution strategy.
We thematically group the scenarios:

\noindent\textbf{Correctness Validation.}
A core set of scenarios to validate the logical correctness of the system: handling various permanent communication exclusions (inbound, outbound, bidirectional, full), managing the lifecycle of temporary disagreements (initiation and expiration), verifying the submodel reuse optimization, and testing the advanced `deep' lifecycle mechanisms like retrospective model rewinding.

\noindent\textbf{Scalability Analysis.}
We conduct scalability experiments that measure growth in computational overhead, \textit{e.g.}, resolution time, client training time, and aggregation time, and the impact on model performance and storage requirements on three dimensions:
\noindent\textbf{Total Client Population}: evaluate how the system scales as the total number of clients increases, while the size and structure of the disagreeing client group remain constant. 
    
\noindent\textbf{Scale of Disagreement Participation}: measure performance as the number of clients actively engaged in disagreements grows within a fixed total client population. This tests the system\textquotesingle{}s response to an increasing proportion of the federation enforcing exclusion rules.
    
\noindent\textbf{Disagreement Density}: We analyze the impact of increasing the number of exclusions enforced by each disagreeing client. This dimension stress-tests the system under conditions of growing fragmentation, from simple one-to-one exclusions to near-total isolation where each client excludes all others.

\noindent\textbf{Execution Environment:} We conduct all experiments on a single machine with Intel Core i7-9750H CPU, 16 GB RAM, and an NVIDIA GeForce GTX 1650 GPU, 4 GB VRAM.

\noindent\textbf{Datasets and Models.} To ensure a comprehensive evaluation, we use two distinct and complementary datasets.

\noindent\textbf{N-CMAPSS} dataset~\cite{Chatterjee2021ncmapss} provides non-IID time-series data simulating the run-to-failure trajectories of a fleet of turbofan engines, thus highly relevant to our motivating use case of predictive maintenance in aviation. The prediction task is Remaining Useful Life.
We use a Multilayer Perceptron model. The model\textquotesingle{}s simplicity is intentional: we isolate and measure the performance of the disagreement resolution mechanism without the confounding factors of a complex model.

\noindent\textbf{MNIST} dataset~\cite{lecun1998gradient} is a widely used image classification benchmark, consisting of handwritten digits. Its ease of use enables rapid and extensive testing of our system\textquotesingle{}s core logic across a large number of complex scenarios, which is computationally prohibitive for N-CMAPSS. We use a CNN.

\begin{figure}
    \centering
    \includegraphics[width=\columnwidth]{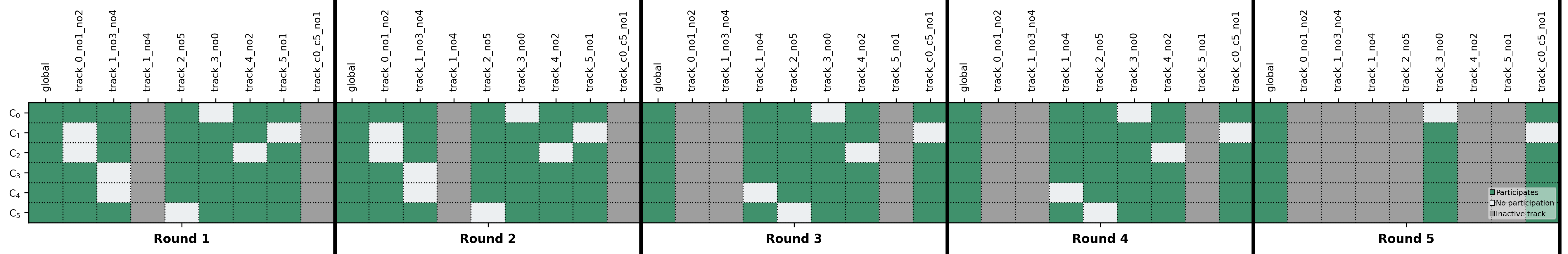}\vspace{-2mm}
    \caption{Track contributions for Scenario 6. The visualization correctly shows the dynamic creation and expiration of multiple tracks in response to overlapping temporary disagreements.}
    \label{fig:s6_track_contributions}
    \vspace{-4mm}
\end{figure}

\vspace{-1mm}
\section{Results}
\label{sec:results}
\vspace{-1mm}


\noindent\textbf{Correctness Validation.} To validate the fundamental correctness of our approach, our tooling automatically confirms the system's behavior across all scenarios. In S1\footnote{ Detailed descriptions and results for all scenarios are available at \url{https://github.com/DaanRosendal/fl-disagreement-resolution}~\cite{rosendal2025fldisagreementresolution}} -- a single permanent inbound exclusion,  the system correctly creates two isolated tracks, with a personalized track for the excluding client and a global track for others, while enforcing all necessary background participations to ensure fairness. Our tooling generates auditable visualizations:  Fig.~\ref{fig:s6_track_contributions} shows the highly dynamic Scenario 6 (S6) where multiple overlapping permanent and temporary disagreements expire at different rounds. The visualization correctly captures the complex, evolving track structure, showing the dynamic creation and dissolution of multiple temporary tracks (e.g., `track\_0\_no1\_no2', `track\_1\_no3\_no4') while permanent tracks persist. 
We validate advanced temporal features through a `deep rewind' scenario, where the system re-aggregates historical models to retroactively enforce an exclusion initiated mid-session. 

\begin{figure}[t]
    \vspace{-5mm}
    \centering
    \begin{subfigure}[b]{\columnwidth}
        \centering
        \includegraphics[width=\textwidth]{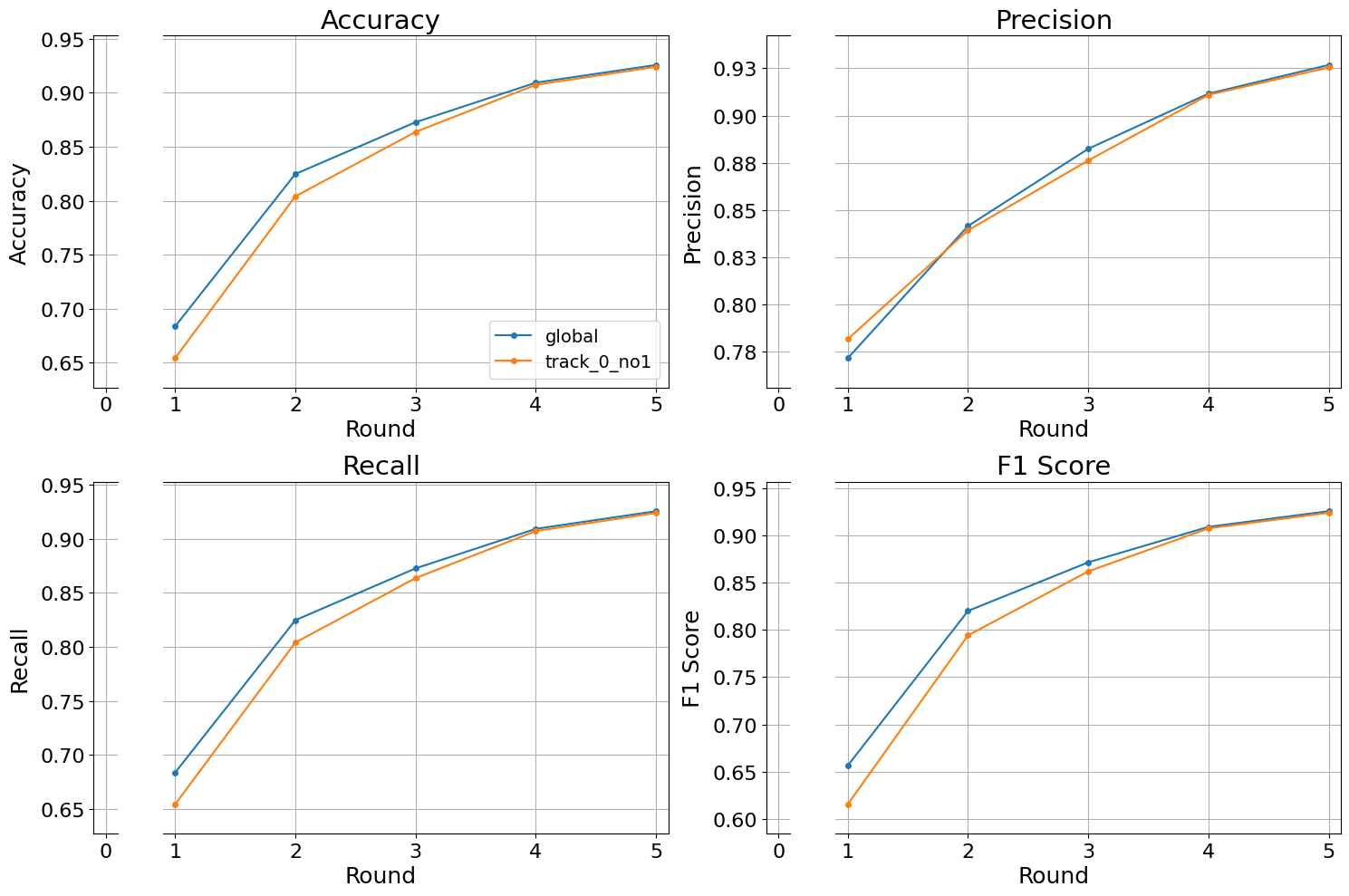}\vspace{-2mm}
        \caption{S1 model performance results for MNIST.}
        \label{fig:s1_mnist}
    \end{subfigure}
    \vfill
    \begin{subfigure}[b]{\columnwidth}
        \centering
        \includegraphics[width=\textwidth]{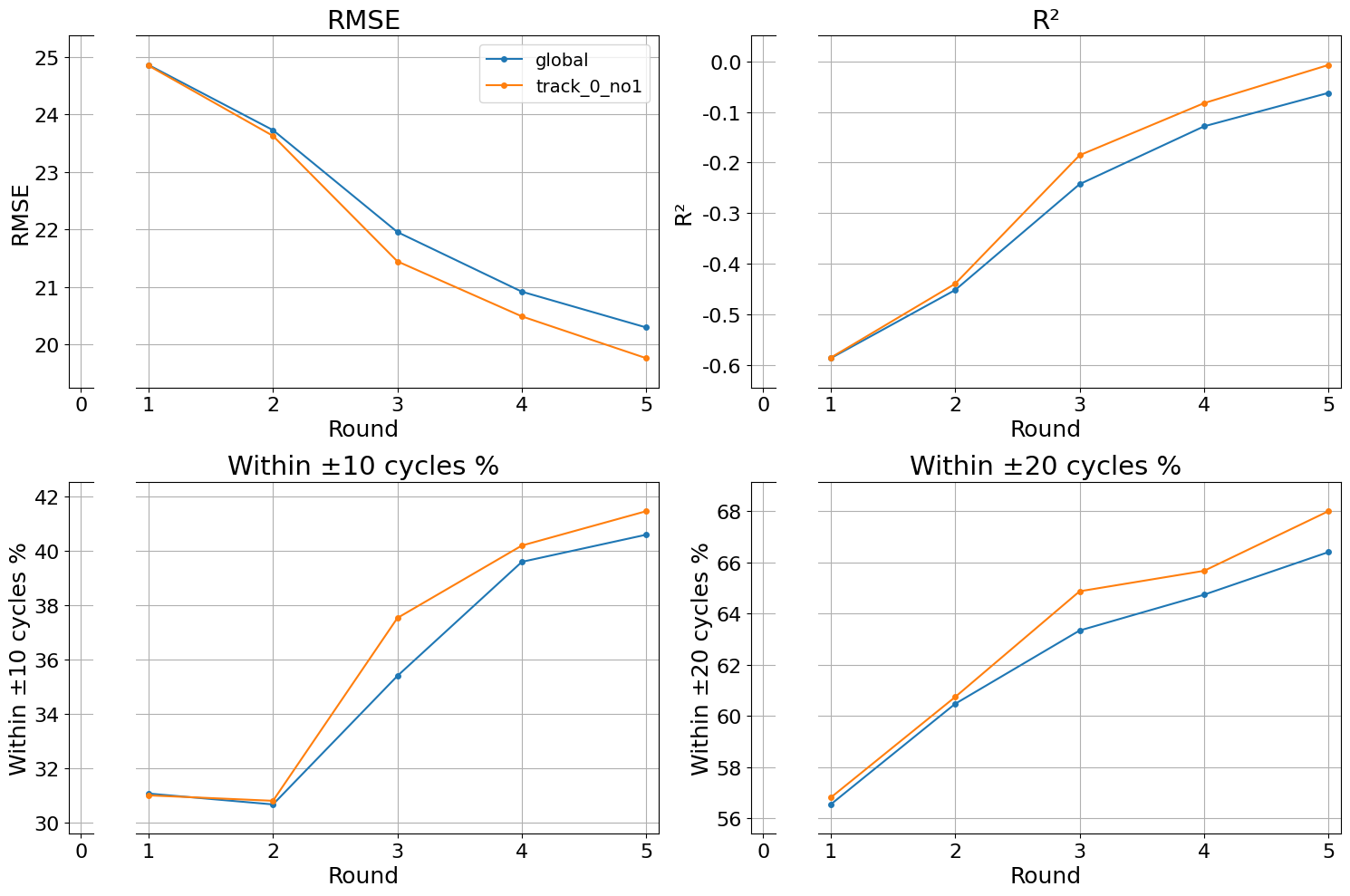}\vspace{-2mm}
        \caption{S1 model performance results for N-CMAPSS.}
        \label{fig:s1_ncmapss}
    \end{subfigure}\vspace{-2mm}
    \caption{Model performance for Scenario 1 (S1) on both datasets. Both tracks demonstrate stable learning, indicating the resolution mechanism does not impede model training.}
    \label{fig:s1_model_performance}
    \vspace{-0.5mm}
\end{figure}

Fig.~\ref{fig:s1_model_performance} shows S1 model performance. 
The isolated personalized track and the main global track achieve effective model training. The stable learning trend holds across all scenarios.

\begin{figure}
    \vspace{-7mm}
  \centering
  \begin{subfigure}[b]{0.49\columnwidth}
    \includegraphics[width=\linewidth]{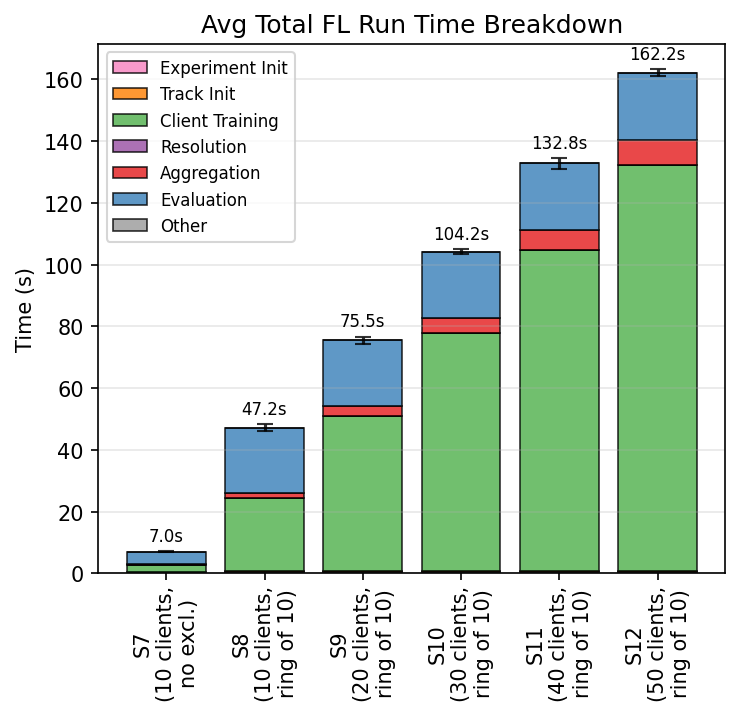}\vspace{-1mm}
    \caption{Lightest scalability scenario set (S7–S12): Total clients increase while disagreements are constant. Total runtime shows linear growth.}
    \label{fig:lightest_scalability}
  \end{subfigure}
  \hfill
  \begin{subfigure}[b]{0.49\columnwidth}
    \includegraphics[width=\linewidth]{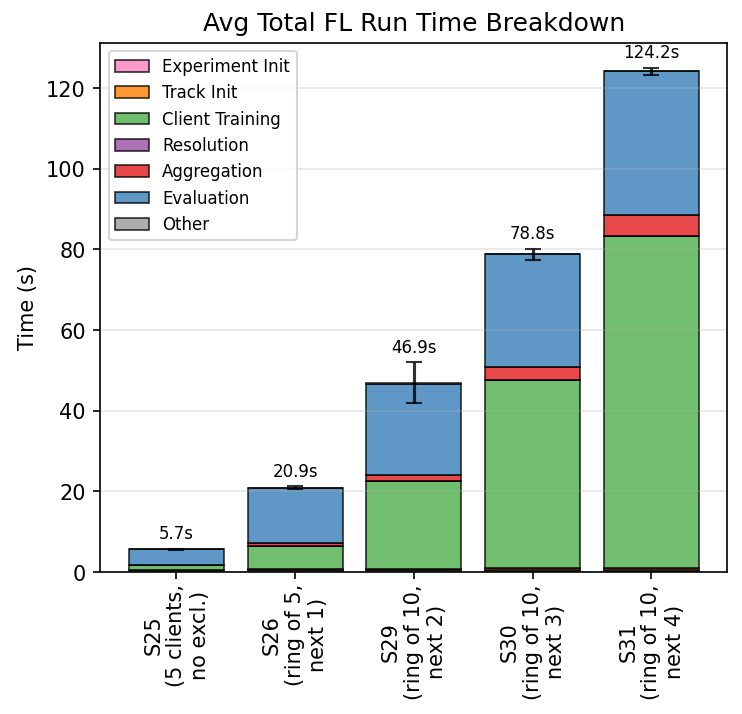}\vspace{-2mm}
    \caption{Heaviest scalability scenario set (S25, S26, S29–S31): Client count, disagreeing participants, and disagreement density increase, with super-linear growth.}
    \label{fig:heaviest_scalability}
  \end{subfigure}
  \caption{Average total FL runtime breakdown for the lightest and heaviest scalability scenarios. Client training time dominates cost; resolution and aggregation overheads are minimal.\vspace{-5mm}}
  \label{fig:scalability_breakdown}
\end{figure}

\noindent\textbf{Scalability Analysis.} \textcolor{black}{We measure the execution time of the core phases (resolution, training, aggregation). Our centralized simulation assumes that excluding network costs allows for an isolated analysis of the core algorithms' computational overhead.} We report the mean performance for scalability results averaged over five distinct runs.
Fig.~\ref{fig:scalability_breakdown} shows the average total FL runtime breakdown per round for two sets of scenarios. Fig.~\ref{fig:lightest_scalability} shows a linear increase in runtime as the number of non-disagreeing clients grows while the disagreement structure is constant. Fig.~\ref{fig:heaviest_scalability} shows a super-linear runtime growth. This is due to testing the most demanding condition where the system scales across three dimensions simultaneously: the total number of clients, the number of clients participating in disagreements, and the disagreement density (the number of exclusions per client).


\vspace{-1mm}
\section{Discussion}
\label{sec:discussion}
\vspace{-1mm}

Our experimental results validate that the proposed multi-track resolution strategy is a correct and scalable method for resolving client-level disagreements in Federated Learning. We discuss the principal findings and their implications below.



\noindent\textbf{RQ1}: We define a formal taxonomy of disagreements, categorizing them into communication-based types, where a client can reject updates from specific peers (inbound), block its own updates from reaching them (outbound), enforce mutual exclusion (bidirectional), or fully withdraw from the process, alongside data-based (partial) exclusions. We define these further by temporal properties: a duration (indefinite or temporary) and a depth, determining if an exclusion is shallow (affecting only future rounds) or deep (requiring retrospective removal of historical influence).

\noindent\textbf{RQ2}: Disagreements are resolved using a robust multi-track strategy that creates isolated model update paths (`tracks') for each unique set of client exclusions. This strategy guarantees strict isolation, preventing the cross-contamination seen in naive approaches where an excluded client\textquotesingle{}s influence leaks back via intermediaries. It ensures fairness by mandating background participation, such that clients\textquotesingle{} updates for any track are generated by training on that specific track\textquotesingle{}s model, maintaining consistent training conditions for all contributions.

\noindent\textbf{RQ3}: We mitigate the combinatorial explosion of tracks scalability challenge in two ways: 1) an efficient server-side resolution algorithm (Alg.~\ref{alg:track_creation}) that is not a system bottleneck (runtime $<1$ ms per round); 2) a submodel reuse optimization consolidates clients with identical exclusion patterns into a single track, reducing model duplication. The main scalability constraint shifts from the server\textquotesingle{}s decision logic to downstream execution workload (client training, aggregation), scaling predictably with the number of unique disagreement patterns.

\noindent\textbf{Insights Beyond the Research Questions.} Our experiments reveal three notable insights: 1) the server-side resolution logic is computationally negligible and does not present a scalability bottleneck -- the primary performance constraints lie in the execution phase of multi-track client training and aggregation; 2) the flexibility of disagreements introduces risks of strategic misuse, such as resource exhaustion or model-differencing privacy attacks, underscoring the need for robust governance mechanisms; 3) submodel reuse proves to be a critical optimization, effectively mitigating track multiplication by consolidating clients with identical exclusion preferences.

\noindent\textbf{Limitations.} The centralized simulation does not account for real-world network effects, \textit{e.g.}, latency.
The scale of our experiments is limited (50 clients) and lacks massive datasets, so it may not fully represent large-scale FL dynamics. Our analysis prioritizes computational overhead over communication payloads, albeit our open-source tooling can generate these metrics. Our unscalable track naming scheme requires a more robust method, \textit{e.g.}, hash-based identifiers.

\noindent\textbf{Future Work.}
We identify three directions for future work: 1) strengthening verification and trust using verifiable unlearning and confidential computing to cryptographically audit policy enforcement, while also evaluating privacy implications; 2) adapting the architecture for real-world deployment by transitioning the system to a distributed middleware, \textit{e.g.}, DYNAMOS. This will enable testing at massive scale under realistic network conditions, additionally allowing for a shift from explicit client-to-client exclusions to policy-based rules for large, attribute-defined groups,\textit{ e.g.}, geographic location; 3) the resolution mechanism can be extended with more granular partial data exclusions~\cite{Routh2024Granularity}, enabling more precise, data-sovereign collaboration, with its value demonstrated through domain-level KPIs, like the predictive maintenance metric.

\section{Conclusion}
\label{sec:conclusion}

We introduced a robust multi-track resolution strategy for resolving client-level disagreements in Federated Learning. We demonstrated that by creating and managing isolated model update paths, it is possible to guarantee strict client exclusion and fairness. Our scalability analysis shows the decision-making logic for translating complex disagreements into a track structure is highly efficient and not a system bottleneck. Instead, the primary scalability constraint is the downstream execution workload of multi-track client training, which scales predictably with the number of unique disagreement patterns. 

By validating the multi-track architecture and its enabling algorithms, our work provides a scalable and architecturally sound method for integrating complex policy compliance directly into the FL lifecycle, ensuring that strategic client autonomy can coexist with effective collaborative learning.

\vspace{-2mm}
\section*{Acknowledgments}\vspace{-1mm}
This work is supported by the project: ``Budget-neutral sustainable home improvements (MOOI-224049)''.
\vspace{-2mm}
\renewcommand*{\bibfont}{\normalfont\small}
\printbibliography

\end{document}